# Observation of spin-motive force in ferrimagnetic GdFeCo alloy films

Shun Fukuda[1], Hiroyuki Awano[1] and Kenji Tanabe[1*]

[1]Toyota Technological Institute, Nagoya, 468-8511, Japan

*electronic mail: tanabe@toyota-ti.ac.jp

**Abstract**

   Non-uniform magnetic structures produce emergent electromagnetic phenomena such as the topological Hall effect and the spin-motive force (SMF). The experimental reports on the SMF, however, are very few and the relationship between the SMF and material parameters is still unclear. In this study, we investigated the SMF in ferrimagnetic GdFeCo alloy films using the spin-torque-induced ferromagnetic resonance method and clarified the relationship. The amplitude of the detected SMF becomes larger than that of the transition metal alloy FeCo by the Gd doping and reaches the maximum near a Gd composition of the boundary between in-plane and perpendicularly magnetized films. According to the analytical calculation, the enhancement is related to the trajectory of the magnetization precession. Moreover, we find that the SMF induced by the magnetic resonance is inversely proportional to the square of the damping constant.

Recently, non-uniform magnetic structures such as magnetic domain wall[1-2], magnetic vortex[3-5], anti-vortex[6-8], and skyrmion[9-13] have attracted much attention in both fundamental and applied physics. The magnetic domain wall and skyrmion, which can be controlled by a current, have been studied toward practical application for future memory devices[1,2,11]. Moreover, the non-uniform magnetic structure induces effective electric and magnetic fields for conduction electrons. The effective magnetic field induced by the skyrmions has been detected as a Hall effect, which is termed the topological Hall effect[12-16]. The effective electric field induced by their dynamics is termed the spin-motive force (SMF)[17-30]. These phenomena are of great interest as the emergent electromagnetism in fundamental physics.

The first observation of the SMF had been reported by Yang and Beach et al. in 2009[21]. They accurately controlled magnetic-field-driven domain wall motion in NiFe alloy wires and detected the SMF induced by the motion using the unique lock-in method. After that, several studies on the SMF were also reported by other groups[22-24,29,30]. The theoretical studies[22,26] reported by Yamane et al. suggest that the SMF induced by the magnetic resonance strongly depends on the damping constant and the SMF induced by the magnetic domain wall motion increases with increasing magnetic anisotropy. In all the experimental studies, however, transition metal alloys such as NiFe alloy[21-24,30] and related oxides such as the magnetite $Fe_3O_4$[29] were used and the relationship between material parameters and the SMF is still unclear. Furthermore, the detected voltages are almost 0.1 - 1 μV and we have few guidelines for the material development owing to the enhancement of the SMF.

Here, we investigate the SMF in GdFeCo alloy films and report the influence of Gd on the SMF. The GdFeCo alloy[31-37] is a ferrimagnetic material, where magnetic moment of FeCo is antiparallel to that of Gd. We can fabricate RE-rich films, where net magnetization is parallel to the moment of the rare-earth metal (Gd), and TM-rich films, where the transition metal (FeCo) mainly contributes net magnetization, by controlling the composition of the alloy. The boundary composition between the RE-rich and the TM-rich alloys is termed the magnetization compensation point (MCP, $x = x_c$). The GdFeCo alloy has large perpendicular magnetic anisotropy near the MCP. Moreover, the Gilbert damping constant in GdFeCo strongly depends on the composition. Therefore, GdFeCo is one of the suitable materials in studies on the relationship between the SMF and some material parameters such as the saturation magnetization, the demagnetizing field, the composition, the magnetic anisotropy, and the damping constant.

The $Si_3N_4$(10 nm)/$Gd_x(Fe_{82}Co_{28})_{1-x}$(16 nm)/Pt(10 nm) strips were fabricated by a magnetron sputtering, electron-beam lithography, and a lift-off process. The GdFeCo alloy layer was deposited by a co-sputtering method from Gd and FeCo targets and the composition of the alloy can be changed by optimization of the cathode power. The composition of all the samples was checked by an energy dispersive X-ray spectrometry. The Pt layer under the GdFeCo layer is used as an electrode for rf currents that induces spin torques and(/or) rf Ampère magnetic fields, which is quite similar to an experimental method of the spin-torque ferromagnetic resonance (ST-FMR). The widths of the strips are changed along a longitudinal direction of the strips and the structures are not rectangular, but trapezoidal.

We measured the SMF by using the unique ST-FMR method proposed by Nagata et al.[29] in Fig. 1(a). When an rf-current is injected into a bilayer of Pt/GdFeCo, a magnetic resonance is excited by the rf spin current and the Ampère field induced by the rf current in the Pt layer[38]. Since the rf current density $j_e$ is inversely proportional to the shrinking width of the strip $w$, the cone angle of the magnetization precession is changed along the longitudinal direction of the strips. Thus, the excited magnetic resonance becomes non-uniform dynamics and is expected to induce the SMF along the longitudinal direction. The SMF was measured by using a lock-in and an amplitude modulation methods. The measured dc voltage involves the SMF and the other contribution that originates from the rectification effect of the anisotropic magnetoresistance and the rf current. Although the rectification effect vanishes when the external magnetic field is parallel to the longitudinal direction of the strips, it appears even if the field slightly differs from the longitudinal direction of the strips. The SMF is detected as a symmetrical component centered at zero magnetic field. On the other hand, the rectification voltage involves both symmetric and anti-symmetric components centered at the resonant magnetic field. In particular, both of the contributions are anti-symmetric at zero magnetic fields[38]. Therefore, we extracted the SMF from the measured dc voltage using the symmetry for the magnetic field.

Figure 1(b) shows the measured typical dc voltage, whose shape is similar to the Lorentz function. We extract the SMF by using a fitting function of

$$V = V_0 + V_1\{S_H(H_0, H_R) + S_H(H_0, -H_R)\} + V_2\{S_H(H_0, H_R) - S_H(H_0, -H_R)\} \quad (1)$$
$$+ V_3\{A_H(H_0, H_R) + A_H(H_0, -H_R)\},$$

where $S_H(H_0, H_R) = \Delta H^2/((H_0 - H_R)^2 + \Delta H^2)$ and $A_H(H_0, H_R) = (H_0 - H_R)\Delta H/((H_0 - H_R)^2 + \Delta H^2)$. $H_0$ is the external magnetic field, $H_R$ is the resonant magnetic field and $\Delta H$ is the full width at half maximum (FWHM). The second term has the same symmetry as the SMF. The sign of the SMF does not depend on the sign of the magnetic field, but is determined by the arrangement of the magnetic precession regions. The third and fourth terms attribute to the rectification effect in Fig. 1(c). The fitting provides information on the resonant frequency, the FWHM, the signal amplitude of the SMF, and the rectification effect.

Figures 2(a-b) show the signal amplitude $V_1$ as a function of the right and left width of the strip. Although the signal of almost 100 nV is detected in the trapezoidal strips, the signal disappears in the rectangular strip, $w_r = w_l = 100$ μm. The sign of the detected signal depends on the shrinking sides of the strip. Note that these tendency cannot be expressed by any rectification effects in ST-FMR studies. Moreover, the SMF is represented as

$$V_1 \propto \frac{1}{w_l^2} - \frac{1}{w_r^2} \quad (2)$$

from the calculation (as shown below). The solid curves in Figs. 2(a-b) indicate the fitting curves of the Eq. (2), which is roughly consistent with the experimental results. Hence, these results reveal that $V_1$ is the SMF.

Figure 2(c) shows the Gd-composition dependence of the SMF $V_1$. The deposited GdFeCo film has the MCP at about $x_c = 0.23$ (as shown below). Hence, it is TM-rich when $x < 0.23$ and RE-rich when $x > 0.23$. The GdFeCo film is in-plane magnetized from $x = 0.0$ to $x = 0.17$ and in $x > 0.27$ owing to the large demagnetizing fields, which is proportional to the saturation magnetization. The alloy films at $x = 0.18, 0.27$, which were perpendicularly magnetized, were measured by controlling the magnetization direction via the large in-plane magnetic field. Except for $x = 0.12$, the detected voltage surprisingly increases with increasing Gd composition as $x < 0.17$, which indicates that the SMF can become larger than that in alloys that consist of only transition metals such as the FeCo alloy via the Gd doping. Moreover, we notice that the SMFs in both the TM-rich and the RE-rich films increase near the Gd compositions of the boundary between in-plane and perpendicularly magnetized films.

To understand the enhancement of the SMF via the Gd doping as $x < 0.17$, we model the motion of the magnetic moment in GdFeCo by assuming that net magnetic moment that consists of Gd and FeCo is a single magnetic moment. The assumption can be reasonable in the case that the Gd composition is far away from an angular momentum compensation point(ACP), which is the composition that the angular momenta of FeCo and Gd cancel out[39]. The used Landau–Lifshitz–Gilbert (LLG) equation is represented as

$$\frac{d\bm{n}}{dt} = -\gamma \bm{n} \times \bm{H}^{\text{eff}} + \alpha \bm{n} \times \frac{d\bm{n}}{dt} + \frac{\hbar}{2e\mu_0 M_s t} J_s (\bm{n} \times \bm{s}) \times \bm{n}. \quad (3)$$

Here, $\bm{n}$ is a normalized magnetization vector, $\gamma$ is the gyromagnetic ratio, $\alpha$ is the Gilbert damping constant, $\mu_0$ is the permeability, $M_s$ is the saturation magnetization of GdFeCo, $\bm{s}$ is the spin magnetic moment and $t$ is the thickness of the GdFeCo layer. The xyz-coordinate direction is defined as the xyz arrows in Fig. 1. $\bm{H}^{\text{eff}} = (h_x \quad -(4\pi M_s N_y + A_y)n_y \quad H_0)^T$, where $h_x$ is the Ampère field induced by the rf current, $N_y$ is the demagnetizing coefficient, $A_y$ is the anisotropic constant, and $H_0$ is the external magnetic field. Since the GdFeCo film is perpendicularly magnetized near the MCP ($x = x_c$), we add only the perpendicular magnetic anisotropy to the LLG equation. When $\alpha \ll 1$, $|n_x|, |n_y| \ll 1$, and $|h_x| \ll |H_0|$,

$$\begin{cases} n_x = \dfrac{\{A_H \omega(\gamma h_x) + S_H \omega_R (\beta J_s)\} + i\{-S_H \omega(\gamma h_x) + A_H \omega_R (\beta J_s)\}}{\alpha \omega_R \gamma (2H_R + k)} \\ n_y = \dfrac{\{-S_H \omega(\gamma h_x) + A_H \omega_R (\beta J_s)\} + i\{-A_H \omega(\gamma h_x) - S_H \omega_R (\beta J_s)\}}{\alpha \omega \gamma (2H_R + k)} \end{cases} \quad (4)$$

where $k = 4\pi M_s N_y + A_y$, $\omega = \gamma \sqrt{H_R(H_R + k)}$ is an angular frequency of the rf current, $\omega_R = \gamma H_R$, and $\beta = \hbar \|\bm{s}\|/2e\mu_0 M_s t$. By substituting $N_y = 1$ and $A_y = -H_y^{ani}$, we can obtain the relationship between the resonant field and the frequency $f$,

$$f = \frac{\gamma}{2\pi}\sqrt{H_R(H_R + 4\pi M_s - H_y^{ani})}. \tag{5}$$

$S_H$ and $A_H$ are the Lorentz and the anti-Lorentz functions in the Eq. (1), respectively. These FWHMs $\Delta H$ are proportional to the frequency,

$$\Delta H = \frac{2\pi\alpha}{\gamma}f. \tag{6}$$

Assuming that $h_x = he^{i\omega t}$ and $J_s = je^{i\omega t}$, we can obtain

$$\begin{cases} n_x = \frac{\sqrt{S_H}\sqrt{(\omega\gamma h)^2 + (\omega_R\beta j)^2}}{\alpha\gamma(2H_R + k)}\frac{1}{\omega_R}\cos(\omega t + \delta) \\ n_y = \frac{\sqrt{S_H}\sqrt{(\omega\gamma h)^2 + (\omega_R\beta j)^2}}{\alpha\gamma(2H_R + k)}\frac{1}{\omega}\sin(\omega t + \delta) \end{cases}, \tag{7}$$

$$\tan\delta = \frac{A_H\omega_R\beta j - S_H\omega\gamma h}{S_H\omega_R\beta j + A_H\omega\gamma h}$$

which reveal that the magnetization trajectory is not a circular orbit, but an elliptical orbit when $4\pi M_s - H_y^{ani} \neq 0$. In particular, its ellipticity mainly depends on the magnetic anisotropy and the demagnetizing field and is independent of the ratio of the rf spin current to the rf Ampère field. On the other hand, the phase of the trajectory is strongly influenced by the ratio of the rf spin current to the rf Ampère field.

Figures 3(a-c) show $H_R$ and $\Delta H$, and $V_1$ as a function of the frequency of the rf current. $4\pi M_s - H_y^{ani}$ is estimated by using the fitting function of the Eq. (5). Since the anisotropy field is parallel to the demagnetizing field, $4\pi M_s - H_y^{ani}$ is regarded as a single parameter. The saturation magnetization $4\pi M_s$ is estimated from the M-H curve in Fig. 3(d). The damping constant $\alpha$ is obtained from a linear fitting of the Eq. (6) including the extrinsic damping constant[40].

The electric field induced by the SMF is represented as

$$E_z = -\frac{\hbar}{2e}\tilde{\mathbf{n}}\cdot\left(\frac{\partial\tilde{\mathbf{n}}}{\partial t}\times\frac{\partial\tilde{\mathbf{n}}}{\partial z}\right), \tag{8}$$

where $\hbar$ is the Plank constant, $e$ is the elementary charge, and $\tilde{\mathbf{n}}$ is the direction of the internal exchange field in a ferromagnetic material[20], which corresponds to the direction of the magnetization, $\mathbf{n}$. The SMF is calculated by $V_{SMF} = \int E_z dz$. By substituting the Eq. (7) into the Eq. (8),

$$V_{SMF} = S_H\frac{\hbar I^2}{4\alpha^2\gamma e t^2}\left(\frac{1}{w_l^2} - \frac{1}{w_r^2}\right)\frac{(H_R + k)(\gamma\tilde{h})^2 + H_R(\beta\tilde{j})^2}{(2H_R + k)^2}, \tag{9}$$

where, $I$ is the rf current, $\tilde{h} = h/j_e$, and $\tilde{j} = j/j_e$ because both $h$ and $j$ are proportional to the current density, $j_e$. The equation reveals that the SMF should be the Lorentz function and be proportional to $w_l^{-2} - w_r^{-2}$. Since the SMF is inversely proportional to $\alpha^2$, we evaluate the SMF as the product of the SMF and the square of the damping constant, $\alpha^2 V_{SMF}$. Besides, the SMF depends on two unknown parameters, which indicates the contributions of the rf Ampère field and the rf spin current, and we could not analyze the experimental data using the equation. Hence, we roughly assume $\gamma \tilde{h} \cong \beta \tilde{j}$ from the obtained fitting parameters of $V_2$ and $V_3$ in the Eq. (1). Then, $\alpha^2 V_{SMF}$ is represented by

$$\alpha^2 V_{SMF} \propto \left\{ \gamma^2 (4\pi M_s - H_y^{ani})^2 + 4\omega^2 \right\}^{-\frac{1}{2}}, \qquad (10)$$

which means that the SMF decreases with increasing frequency and increases with decreasing $4\pi M_s - H_y^{ani}$. When the equation is applied to the frequency dependence of the SMF in Fig. 3(c), the fitting curve is good agreement with the decreasing tendency in Fig. 3(c), which suggests the validity of the approximation.

Figures 4(a-c) show the Gd-composition dependences of $\alpha$, $4\pi M_s$, $4\pi M_s - H_y^{ani}$, and $\alpha^2 V_1$. The dependence of $4\pi M_s$, which is obtained from the M-H curve, shows that the MCP is near $x = 0.23$ and $4\pi M_s - H_y^{ani}$ decreases with increasing Gd-composition toward the boundary composition between the in-plane and perpendicularly magnetized films. The damping constant trends to increase near the MCP, which is also consistent with the previous report[32,39]. The theoretical suggestion[32] indicates that the damping constant diversifies to infinity at the ACP, which is near the MCP owing to slight differences between Fe, Co, and Gd g-factors. Though the ACP is measured by the Barnett effect measurement technique[40,41], we estimate the ACP at $x = 0.21$ using g factors $g_{Gd} = 2.0$ and $g_{FeCo} = 2.2$. The SMF is evaluated as $\alpha^2 V_1$ in Fig. 4(c) to take into account the influence of the damping constant. As a results, $\alpha^2 V_1$ increases with increasing Gd composition up to $x = 0.16$, which is close to the boundary composition, and decreases with increasing Gd composition after that. The results reveal that the suppression of $V_1$ at $x = 0.12$ in Fig. 2(c) stems from the apparent effect due to the increase in the damping constant. The curve in Fig. 4(c), which is a guide by eye, is approximately drawn by using the Eq. (10). Since the curve overlaps with the experimental data, the enhancement via the Gd doping can be roughly expressed by the competition between the demagnetizing field and the magnetic anisotropy field. In other words, when the trajectory of the magnetization precession becomes circular, the output of the SMF is maximized.

We have investigated the SMF induced in the ferrimagnetic GdFeCo alloy films by using the unique spin-torque ferromagnetic resonance method. The amplitude of the SMF, which strongly depends on the Gd composition, is maximized at the boundary composition between the in-plane and the perpendicularly magnetized films, which shows that the trajectory of the magnetization precession is crucial in order to enhance the SMF. Since the SMF induced by the magnetic resonance is inversely proportional to the square of the damping constant, ferromagnetic materials with lower damping constants are desired to enhance the SMF.

We would like to thank Mr. A. Takahashi and Dr. S. Sumi for collaboration at an early stage of this work. This work was partly supported by a Grant-in-Aid for Young Scientists (B) (No. 18K14123) from JSPS and a Toyota Riken scholar from the Toyota Physical and Chemical Research Institute.

**References**

1)  A. Yamaguchi, T. Ono, S. Nasu, K. Miyake, K. Mibu, and T. Shinjo, Phys. Rev. Lett. **92**, 077205 (2004).

2)  S. S. P. Parkin, M. Hayashi, and L. Thomas, Science **320**, 190 (2008).

3)  K. Yamada, S. Kasai, Y. Nakatani, K. Kobayashi, H. Kohno, A. Thiaville, and T. Ono, Nat. Mater. **6**, 270 (2007).

4)  K. Nakano, K. Tanabe, R. Hiramatsu, D. Chiba, N. Ohshima, S. Kasai, T. Sato, Y. Nakatani, K. Sekiguchi, K. Kobayashi, and T. Ono, Appl. Phys. Lett. **102**, 072405 (2013).

5)  K. Tanabe, D. Chiba, and T. Ono, Jpn. J. Appl. Phys., Part 1 **49**, 078001 (2010).

6)  K. Shigeto, T. Okuno, K. Mibu, T. Shinjo, and T. Ono, Appl. Phys. Lett. **80**, 4190 (2002).

7)  K. Tanabe and K. Yamada, Appl. Phys. Lett. **110**, 132405 (2017).

8)  K. Tanabe and K. Yamada, Appl. Phys. Express **11**, 113003 (2018).

9)  U. K. Rössler, A. N. Bogdanov, and C. Pfleiderer, Nature **442**, 797 (2006).

10) S. Mühlbauer, B. Binz, F. Jonietz, C. Pfleiderer, A. Rosch, A. Neubauer, R. Georgii, and P. Böni, Science **323**, 915 (2009).

11) A. Fert, V. Cros, and J. Sampaio, Nat. Nanotechnol. **8**, 152 (2013).

12) M. Lee, W. Kang, Y. Onose, Y. Tokura, and N. P. Ong, Phys. Rev. Lett. **102**, 186601 (2009).

13) A. Neubauer, C. Pfleiderer, B. Binz, A. Rosch, R. Ritz, P. G. Niklowitz, and P. Böni, Phys. Rev. Lett. **102**, 186602 (2009).

14) J. Ye, Y. B. Kim, A. J. Millis, B. I. Shraiman, P. Majumdar, and Z. Tešanović, Phys. Rev. Lett. **83**, 3737 (1999).

15) Y. Taguchi, Y. Oohara, H. Yoshizawa, N. Nagaosa, and Y. Tokura, Science **291**, 2573 (2001).

16) Y. Machida, S. Nakatsuji, S. Onoda, T. Tayama, and T. Sakakibara, Nature **463**, 210 (2010).

17) L. Berger, Phys. Rev. B **33**, 1572 (1986).

18) G. E. Volovik, J. Phys. C **20** L83 (1987).

19) A. Stern, Phys. Rev. Lett. **68**, 1022 (1992).

20) S. E. Barnes and S. Maekawa, Phys. Rev. Lett. **98**, 246601 (2007).

21) S. A. Yang, G. S. D. Beach, C. Knutson, D. Xiao, Q. Niu, M. Tsoi, and J. L. Erskine, Phys. Rev. Lett. **102**, 067201 (2009).

22) Y. Yamane, K. Sasage, T. An, K. Harii, J. Ohe, J. Ieda, S. E. Barnes, E. Saitoh, and S. Maekawa, Phys. Rev. Lett. **107**, 236602 (2011).

23) K. Tanabe, D. Chiba, J. Ohe, S. Kasai, H. Kohno, S. E. Barnes, S. Maekawa, K. Kobayashi, and T. Ono, Nat. Commun. **3**, 845 (2012).

24) M. Hayashi, J. Ieda, Y. Yamane, J. Ohe, Y. K. Takahashi, S. Mitani, and S. Maekawa, Phys. Rev. Lett. **108**, 147202 (2012).

25) K.-W. Kim, J.-H. Moon, K.-J. Lee, and H.-W. Lee, Phys. Rev. Lett. **108**, 217202 (2012).

26) Y. Yamane J. Ieda, and S. Maekawa, Appl. Phys. Lett. **100**, 162401 (2012).

27) J.-i. Ohe, and Y. Shimada, Appl. Phys. Lett. **103**, 242403 (2013).

28) Y. Yamane, S. Hemmatiyan, J. Ieda, S. Maekawa and J. Sinova, Sci. Rep. **4**, 6901 (2014)

29) M. Nagata, T. Moriyama, K. Tanabe, K. Tanaka, D. Chiba, J. Ohe, Y. Hisamatsu, T. Niizeki, H. Yanagihara, E. Kita, and T. Ono, Appl. Phys. Express **8**, 123001 (2015).

30) W. Zhou, T. Seki, H. Imamura, J. Ieda, and K. Takanashi, Phys. Rev. B **100**, 094424 (2019).

31) P. Hansen, C. Clausen, G. Much, M. Rosenkranz, and K. Witter, J. Appl. Phys. **66**, 756 (1989).

32) C. D. Stanciu, A. V. Kimel, F. Hansteen, A. Tsukamoto, A. Itoh, A. Kirilyuk, and T. Rasing, Phys. Rev. B **73**, 220402(R) (2006).

33) M. Binder, A. Weber, O. Mosendz, G. Woltersdorf, M. Izquierdo, I. Neudecker, J. R. Dahn, T. D. Hatchard, J. U. Thiele, C. H. Back, and M. R. Scheinfein, Phys. Rev. B **74**, 134404 (2006).

34) X. Jiang, L. Gao, J. Z. Sun, and S. S. P. Parkin, Phys. Rev. Lett. **97**, 217202 (2006).

35) K. J. Kim, S. K. Kim, Y. Hirata, S. H. Oh, T. Tono, D. H. Kim, T. Okuno, W. S. Ham, S. Kim, G. Go, Y. Tserkovnyak, A. Tsukamoto, T. Moriyama, K. J. Lee, and T. Ono, Nat. Mater. **16**, 1187 (2017).

36) D.-H. Kim, T. Okuno, S. K. Kim, S.-H. Oh, T. Nishimura, Y. Hirata, Y. Futakawa, H. Yoshikawa, A. Tsukamoto, Y. Tserkovnyak, Y. Shiota, T. Moriyama, K.-J. Kim, K.-J. Lee, and T. Ono, Phys. Rev. Lett. **122**, 127203 (2019).

37) T. Seki, A. Miura, K. Uchida, T. Kubota, and K. Takanashi, Appl. Phys. Express **12**, 023006 (2019).

38) L. Liu, T. Moriyama, D. C. Ralph, and R. A. Buhrman, Phys. Rev. Lett. **106**, 036601 (2011).

39) T. Okuno, S. K. Kim, T. Moriyama, D.-H. Kim, H. Mizuno, T. Ikebuchi, Y. Hirata, H. Yoshikawa, A. Tsukamoto, K.-J. Kim, Y. Shiota, K.-J. Lee and T. Ono, Applied Physics Express **12**, 093001 (2019).

40) M. A. W. Schoen, D. Thonig, M. L. Schneider, T. J. Silva, H. T. Nembach, O. Eriksson, O. Karis and J. M. Shaw, Nature Physics **12**, 839 (2016).

41) M. Imai, Y. Ogata, H. Chudo, M. Ono, K. Harii, M. Matsuo, Y. Ohnuma, S. Maekawa, and E. Saitoh, Appl. Phys. Lett. **113**, 052402 (2018).

42) M. Imai, H. Chudo, M. Ono, K. Harii, M. Matsuo, Y. Ohnuma, S. Maekawa, E. Saitoh, Appl. Phys. Lett. **114**, 162402 (2019).

**Figure 1**

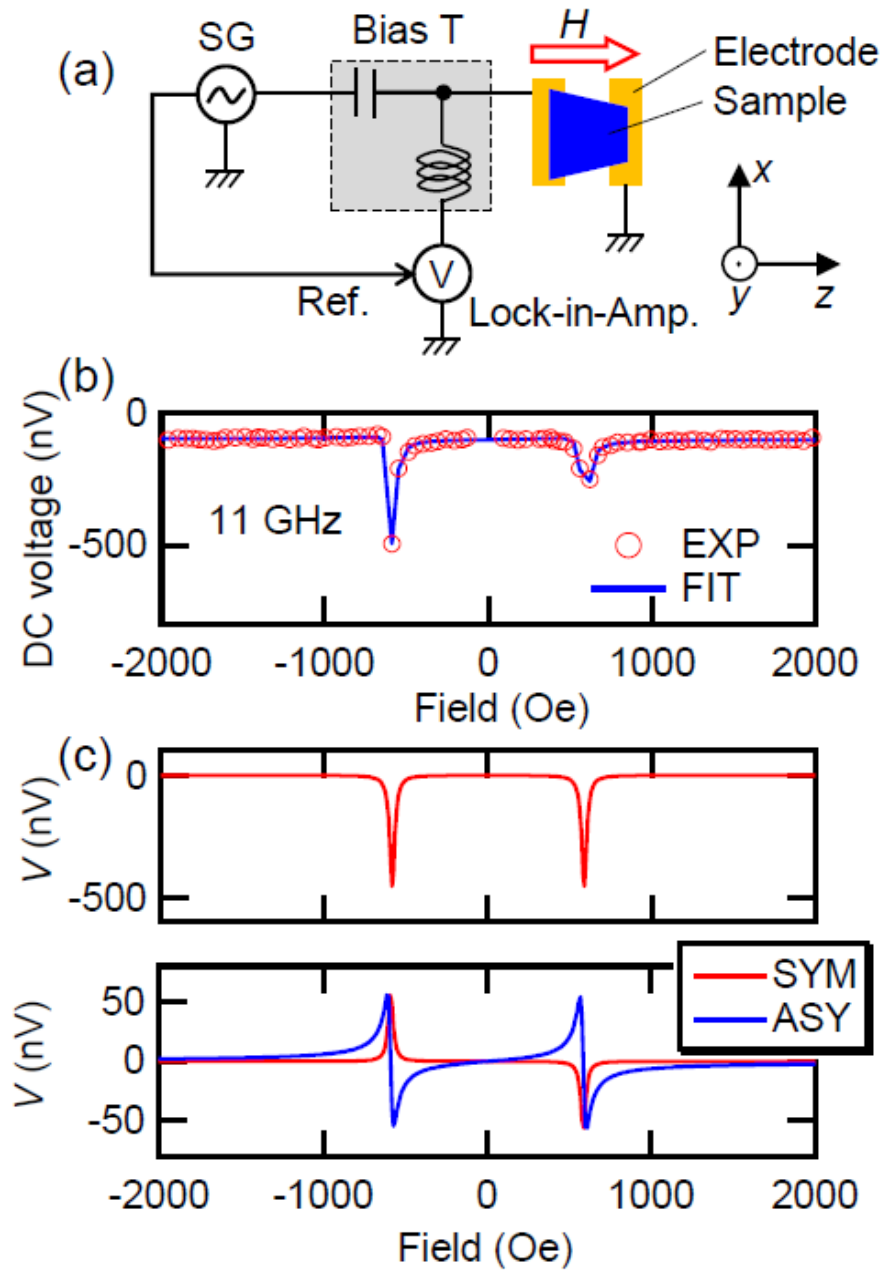

Fig. 1. (a) Schematic illustration of the measurement setup. (b) The detected typical dc voltage as a function of the external magnetic field. The blue curve represents the fitted data. The power of the rf current is 10 dBm. The $Fe_{82}Co_{18}$ alloy is used. (c) Each component of the fitting function. The data in the upper figure have the same symmetry as the SMF. On the other hand, the data in the lower figure have the same symmetry as the signals of the rectification effect.

**Figure 2**

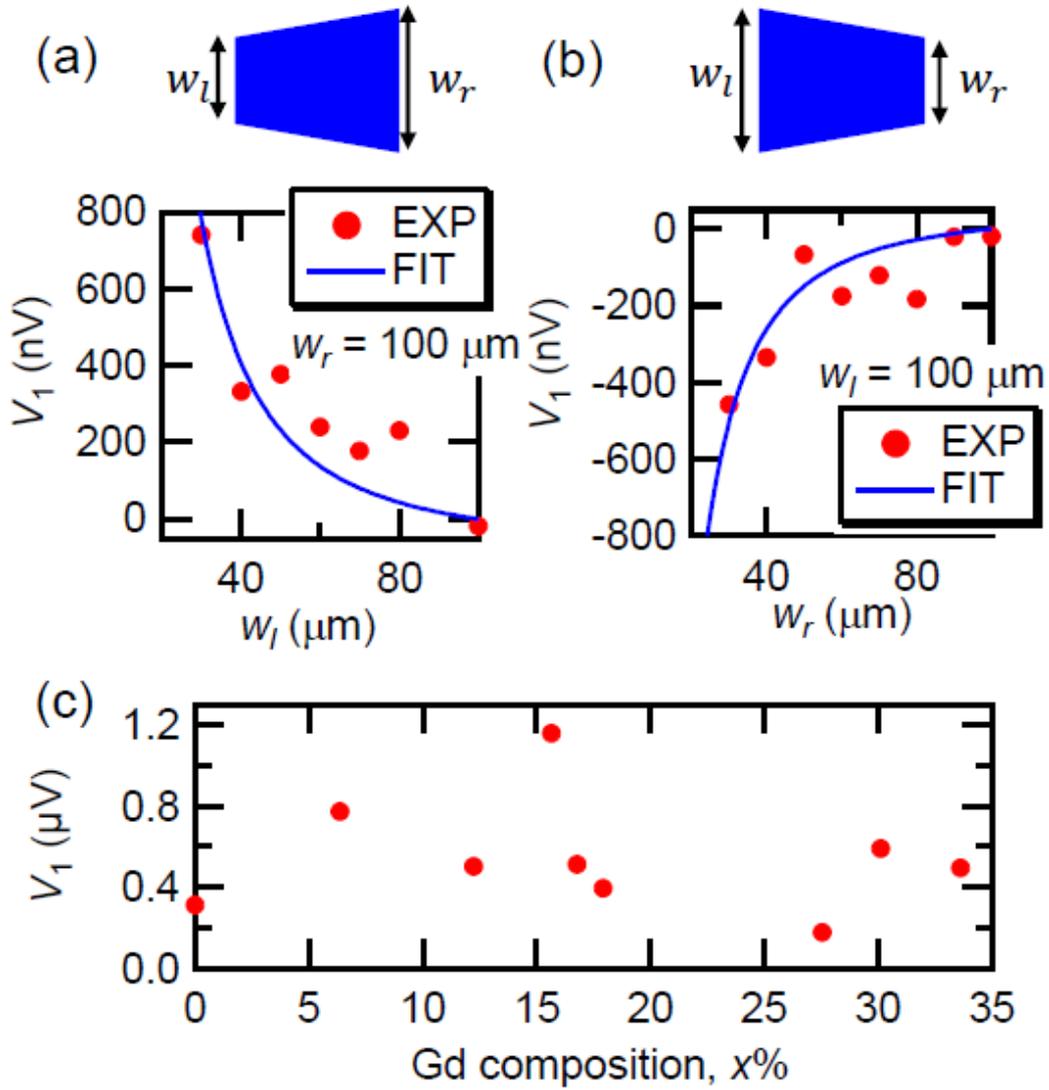

Fig. 2. (a-b) The detected voltages, $V_1$, as a function of the widths of the left and the right edges of the ferromagnetic strips. The frequency and the power of the rf current are 6 GHz and 10 dBm, respectively. $w_r$ and $w_l$ are defined in the insets. (c) The Gd composition dependence of the detected voltages, $V_1$. The frequency and the power of the rf current are 6 GHz and 10 dBm, respectively. The width of the left edge is 100 μm and the width of the right edge is 20 μm.

**Figure 3**

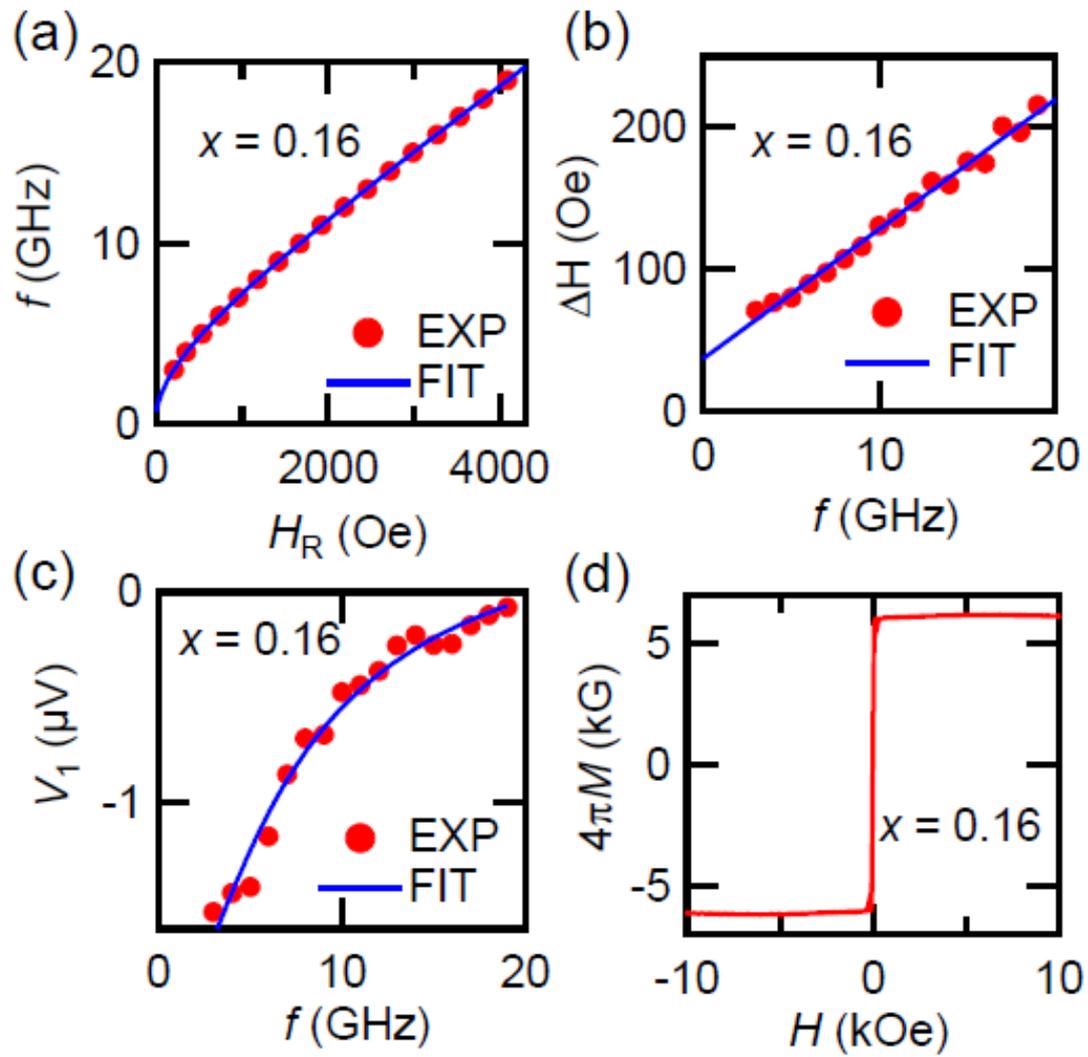

Fig. 3. (a-c) The resonant field, the FWHM, and the SMF as a function of the frequency of the rf current. The power of the rf current is 10 dBm. The width of the left edge is 100 μm and the width of the right edge is 20 μm. (d) The magnetization-field curve. The composition of the used alloy is $x = 0.16$. The magnetic field direction is in-plane.

**Figure 4**

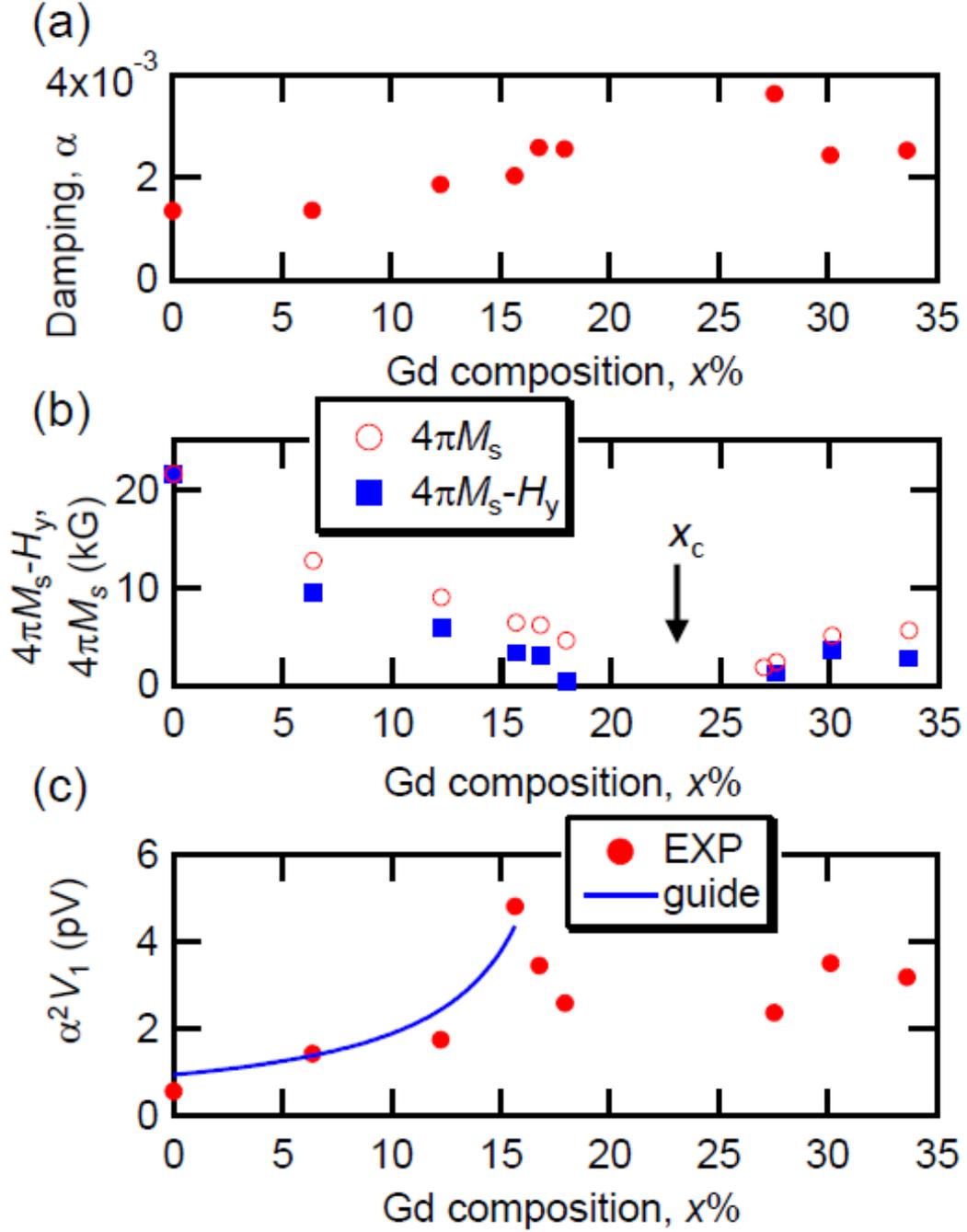

Fig. 4. (a) The Gilbert damping constant as a function of the Gd composition. (b) The saturation magnetization $4\pi M_s$ and $4\pi M_s - H_y$ as a function of the Gd composition. (c) $\alpha^2 V_1$ as a function of the Gd composition. The frequency and the power of the rf current are 6 GHz and 10 dBm, respectively. The blue curve is a guide for the eye, which is written by using the Eq. (10).